%% file: connolly.tex
\newcommand{\boldthing}[1]{\mbox{\bf #1}}
\newcommand{\myindex}[1]{\mbox{\scriptsize #1}}
\newcommand{\aij}{a_{ij}}
\newcommand{\aik}{a_{ik}}
\newcommand{\asupsub}[2]{a^{\myindex{#1}}_{#2}}
\newcommand{\ajmax}{\asupsub{max}{j}}
\newcommand{\ajmin}{\asupsub{min}{j}}
\newcommand{\akmax}{\asupsub{max}{k}}
\newcommand{\assign}{\leftarrow}
\newcommand{\boldx}{\boldthing{x}}
\newcommand{\classsub}[1]{c_{#1}}
\newcommand{\classone}{\classsub{1}}

\newcommand{\classj}{\classsub{j}}
\newcommand{\classk}{\classsub{k}}
\newcommand{\cov}{\mbox{$\boldmath \Sigma$}}
\newcommand{\covsub}[1]{{\cov}_{#1}}

\newcommand{\covj}{\covsub{j}}

\newcommand{\half}{\frac{\mbox{\scriptsize 1}}{\mbox{\scriptsize{2}}}}

\newcommand{\kdtree}{{\em k}d-tree}
\newcommand{\makestats}{\mbox{\sc MakeStats}}
\newcommand{\mbw}{\mbox{\em MBW}}
\newcommand{\mhdist}{\mbox{\em MHD}}
\newcommand{\mhdsup}[1]{\mhdist^{\myindex{#1}}}
\newcommand{\mhdmax}{\mhdsup{max}}
\newcommand{\mhdmin}{\mhdsup{min}}
\newcommand{\model}{\mbox{$\boldmath \theta$}}
\newcommand{\modelsup}[1]{{\model}^{\myindex{#1}}}
\newcommand{\modelt}{\modelsup{t}}
\newcommand{\mrkdtree}{{\em mrk}d-tree}
\newcommand{\mrkdtrees}{{\em mrk}d-trees}
\newcommand{\boldmu}{\boldmath \mu}
\newcommand{\musub}[1]{{\boldmu}_{#1}}

\newcommand{\muj}{\musub{j}}
\newcommand{\node}{\mbox{\footnotesize \sc Nd}}
\newcommand{\nodesub}[1]{\mbox{{\node}.{\footnotesize \sc #1}}}
\newcommand{\nodecentroid}{\nodesub{centroid}}
\newcommand{\nodecov}{\nodesub{cov}}
\newcommand{\nodehrect}{\nodesub{hyperrect}}
\newcommand{\nodeleft}{\nodesub{left}}
\newcommand{\nodenumpoints}{\nodesub{numpoints}}
\newcommand{\noderight}{\nodesub{right}}
\newcommand{\nodesplitdim}{\nodesub{splitdim}}
\newcommand{\nodesplitval}{\nodesub{splitval}}
\newcommand{\psub}[1]{p_{#1}}

\newcommand{\pj}{\psub{j}}
\newcommand{\pk}{\psub{k}}
\newcommand{\rootnode}{\mbox{\sc Root}}

\newcommand{\sw}{\mbox{\sc sw}}
\newcommand{\swsub}[1]{{\sw}_{#1}}
\newcommand{\swone}{\swsub{1}}
\newcommand{\swtwo}{\swsub{2}}
\newcommand{\swN}{\swsub{N}}
\newcommand{\swj}{\swsub{j}}
\newcommand{\swx}{\mbox{\sc swx}}
\newcommand{\swxsub}[1]{{\swx}_{#1}}
\newcommand{\swxone}{\swxsub{1}}
\newcommand{\swxN}{\swxsub{N}}
\newcommand{\swxj}{\swxsub{j}}
\newcommand{\swxx}{\mbox{\sc swxx}}
\newcommand{\swxxsub}[1]{{\swxx}_{#1}}
\newcommand{\swxxone}{\swxxsub{1}}
\newcommand{\swxxN}{\swxxsub{N}}
\newcommand{\swxxj}{\swxxsub{j}}
\newcommand{\wsub}[1]{w_{#1}}

\newcommand{\wthreetwo}{\wsub{32}}

\newcommand{\wbarj}{{\bar{w}}_j}
\newcommand{\wij}{\wsub{ij}}
\newcommand{\wsubsup}[2]{w^{\myindex{#2}}_{#1}}
\newcommand{\wjmax}{\wsubsup{j}{max}}
\newcommand{\wjmin}{\wsubsup{j}{min}}
\newcommand{\wkmin}{\wsubsup{k}{min}}
\newcommand{\wjsofar}{\wsubsup{j}{sofar}}
\newcommand{\wjtotal}{\wsubsup{j}{total}}
\newcommand{\xbar}{\bar{\boldx}}
\newcommand{\xsub}[1]{{\boldx}_{#1}}
\newcommand{\xsubi}{\xsub{i}}

\newcommand{\gaussfactor}{((2 \pi)^M \| \covj \|)^{-1/2}}
\newcommand{\commaspace}{\mbox{\ \ \ ,\ \ \ }}
\newcommand{\xsubiinnode}{\mbox{\footnotesize ${\mbox{\bf x}}_i \in \mbox{\sc nd}$}}

\newcommand{\Mpc}{h$^{-1}$ Mpc}

\documentstyle[12pt,aaspp4,psfig]{article}

\begin{document}

\title{Fast Algorithms and Efficient Statistics: Density Estimation in 
Large Astronomical Datasets}

\author{A. J. Connolly\footnote{Authors' names are given in alphabetical order.},} 
\affil{
Department of Physics \& Astronomy, University of Pittsburgh} 
\authoremail{ajc@phyast.pitt.edu}

\author{C. Genovese,} 
\affil{
Department of Statistics, Carnegie Mellon University, 5000 Forbes Avenue,
Pittsburgh, PA-15213} 
\authoremail{genovese@stat.cmu.edu}

\author{A. W. Moore,} 
\affil{Robotics Institute \& the Computer Science
Department, Carnegie Mellon University, 5000 Forbes Avenue,
Pittsburgh, PA-15213} 
\authoremail{awm@cs.cmu.edu}

\author{R. C. Nichol,} 
\affil{Department of Physics, Carnegie Mellon University,
5000 Forbes Avenue, Pittsburgh, PA-15213}
\authoremail{nichol@cmu.edu}

\author{J. Schneider,} 
\affil{Robotics Institute \& the Computer Science
Department, Carnegie Mellon University, 5000 Forbes Avenue,
Pittsburgh, PA-15213} 
\authoremail{schneide@cs.cmu.edu}

\author{L. Wasserman,} 
\affil{
Department of Statistics, Carnegie Mellon University, 5000 Forbes Avenue,
Pittsburgh, PA-15213} 
\authoremail{larry@stat.cmu.edu}

\begin{abstract}
In this paper, we outline the use of Mixture Models in density estimation of
large astronomical databases. This method of density estimation has been known
in Statistics for some time but has not been implemented because of the large
computational cost. Herein, we detail an implementation of the Mixture Model
density estimation based on multi--resolutional KD--trees which makes this
statistical technique into a computationally tractable problem. We provide the
theoretical and experimental background for using a mixture model of Gaussians
based on the Expectation Maximization (EM) Algorithm. Applying these analyses
to simulated data sets we show that the EM algorithm -- using the AIC
penalized likelihood to score the fit -- out-performs the best kernel density
estimate of the distribution while requiring no ``fine--tuning'' of the input
algorithm parameters. We find that EM can accurately recover the underlying
density distribution from point processes thus providing an efficient adaptive
smoothing method for astronomical source catalogs. To demonstrate the general
application of this statistic to astrophysical problems we consider two cases
of density estimation: the clustering of galaxies in redshift space and the
clustering of stars in color space. From these data we show that EM provides
an adaptive smoothing of the distribution of galaxies in redshift space
(describing accurately both the small and large-scale features within the
data) and a means of identifying outliers in multi--dimensional color--color
space (e.g.\ for the identification of high redshift QSOs). Automated tools
such as those based on the EM algorithm will be needed in the analysis of the
next generation of astronomical catalogs (2MASS, FIRST, PLANCK, SDSS) and
ultimately in in the development of the National Virtual Observatory.

\end{abstract}


\section{Introduction}

With recent technological advances in wide field survey astronomy it
has now become possible to map the distribution of galaxies and stars
within the local and distant Universe across a wide full spectral
range (from X-rays through to the radio) {\it e.g.}  FIRST, MAP,
Chandra, HST, ROSAT, 2dF, Planck. In isolation the scientific returns
from these new data sets will be enormous; combined these data will
represent a new paradigm for how we undertake astrophysical
research. They present the opportunity to create a ``Virtual
Observatory'' (Szalay and Brunner 1999) that will enable a user to
seamlessly analyze and interact with a multi-frequency digital map of
the sky.

A prime driver of the ``Virtual observatory'', and an example of the
challenges these new data sets will provide, is the Sloan Digital Sky
Survey (York et al 2000). This multicolor survey will map 10,000 sq
degrees centered at the North Galactic Cap to a depth of $g<23.5$. It
will result in the detection of over 200 million stars, galaxies and
QSOs. For each of these sources over 500 individual attributes will be
measured (positions, sizes, colors, profiles and morphologies)
resulting in a catalog that is likely to exceed 1 Terabyte in
size. While the wealth of information contained within this survey is
clear the questions we must address is how to we effectively and
efficiently analyze these massive and intrinsically multidimensional
data sets.

Standard statistical approaches do not easily scale to the regime of
10$^8$ data points and 100's of dimensions. We must therefore begin to
explore new efficient and robust methodologies that enable us to carry
out our scientific analyses. In the first of a series of papers we try
and address this issue through the combination of new research in both
statistics and computer science and explore its application to
astronomical datasets. This paper addresses the general problem of
density estimation in astrophysics. This class of problems arises in a
number of different astrophysical applications from identifying
overdensities in the spatial distribution of galaxies (e.g.\ cluster
detection algorithms) through to mapping the density distribution of
the colors of stars in order to identify anomalous records
(e.g. identifying QSO in color-color diagrams).

Most current techniques rely on the application of kernels
with a fixed smoothing paper
(i.e.\ the data are convolved with a kernel of a fixed size that, it
is hoped, has some physical meaning). The resultant density
distributions are then clipped above some heuristically chosen
threshold and those regions above this threshold are identified as
overdensities. A major concern with using such methods is
that the results of the analysis strongly 
depend on the choice of smoothing parameter.
In particular, a fixed smoothing parameter
may tend to oversmooth regions with sharp features
while undersmoothing other regions.
Fixed kernel smoothing, therefore, is not optimized for the full range
of overdensities in astronomical datasets and an adaptive, or
multi--resolution, approach is need to probe both compact and
extended structures at the same time.

We, therefore, seek a mechanism for defining a non-parametric and
compact representation of the underlying density, one
that adapts to the inherently multi-resolution nature of the data.  In
this paper we develop the formalism for just such an adaptive
filtering of multidimensional data using mixture--models. 
Alternative adaptive methods, such as wavelets
(Donoho, Johnstone, Kerkyacharian, and Picard, 1996)
will be discussed in a future paper.
Mixture models
are not new and has been known in statistics for some time.
It has always, however, been considered too difficult to implement
these models for massive datasets
because of the computational intensive nature of the algorithm.  In
Section 3 we discuss multi--resolution KD--trees which is
a data structure that increases the computational speed of these
mixture models by over three orders of magnitude, thus facilitating
their practical use. In Section 4, we present results from the
application of these new methods to the clustering of objects in
redshift and color space.

\begin{center}
\section{A Statistical Analysis of Clustering Using Mixture Models}
\end{center}

There are many statistical methods for estimating densities of
distributions.  In this paper we focus on an adaptive, non-parametric
process namely mixture models (McLachlan and Basford, 1987). We
describe below several aspects of this approach. These are: (i) the
definition of the mixture model, (ii) maximum likelihood estimation
for estimating the parameters of the mixture model, (iii) the EM
algorithm (Dempster, Laird and Rubin, 1977) for finding the maximum
likelihood estimate, (iv) methods for choosing the number of
components in the mixture, (v) measurement error, (vi) comparison to
kernel methods.  Implementing (iii) efficiently is discussed in
Section 3.

\subsection{The Statistical Model}
Let $X^n = (X_1, \ldots , X_n)$ represent the data.  Each $X_i$ is a
$d$-dimensional vector giving, for example, the location of the $i^{th}$
galaxy.  We assume that $X_i$ takes values in a set $A$ where $A$ is a patch
of the sky from which we have observed.  We regard $X_i$ as having
been drawn from a probability distribution
with probability density function $f$.  This means that $f\geq 0$, $\int_A
f(x) dx =1$ and the proportion of galaxies in a region $B$ is given by
$$
Pr (X_i \in B) = \int_B f(x) dx.
$$
In other words, $f$ is the the normalized galaxy density and the
proportion of galaxies in a region $B$ is just the integral of $f$
over $B$.  Our goal is to estimate $f$ from the observed data $X^n$.

The density $f$ is assumed to be of the form
\begin{equation}\label{eq:mix}
f(x;\theta_k) = p_0 U(x) +\sum_{j=1}^k p_j \phi(x ; \mu_j, \Sigma_j) 
\end{equation}
where $\phi(x ; \mu, \Sigma)$ denotes a $d$-dimensional Gaussian with
mean $\mu$ and covariance $\Sigma$:
$$
\phi(x ; \mu, \Sigma)=
\frac{1}{ (2\pi)^{d/2} |\Sigma|^{1/2}}
\exp\left\{ -\frac{1}{2} (x-\mu)^T \Sigma^{-1} (x-\mu)\right\},
$$
and $U(\cdot )$ is a uniform density over $A$ i.e.\  $U(x) = 1/V$ for
all $x\in A$, where $V$ is the volume of $A$.  The unknown parameters
in this model are $k$ (the number of Gaussians) and $\theta_k = (p,
\mu, \Sigma)$ where $p=(p_0, \ldots , p_k)$, $\mu = (\mu_1, \ldots ,
\mu_k)$ and $\Sigma = (\Sigma_1,\ldots , \Sigma_k)$.  Here, $p_j \geq
0$ for all $j$ and $\sum_{j=0}^k p_j=1$.  This model is called a
mixture of Gaussians (with a uniform component).  Intuitively, we can
think of this distribution in the following way: we draw a random
number $G_i$ such that $Pr (G_i =j)=p_j$, $j=0,1,\ldots , k$. If $G_i
=0$ then $X_i$ is drawn from a uniform distribution over $A$. If
$G_i=j$, $j>0$, then we draw $X_i$ from a Gaussian distribution with
mean $\mu_j$ and variance $\Sigma_j$.  We let $\Theta_k$ denote the
set of all possible values of $\theta_k$.  The set $\Theta_k$ is
called the parameter space.  (Technically, the Gaussians should be
truncated and normalized over $A$ but this is a minor point which we
ignore.)

The uniform term represents ``clutter,'' or background field,
i.e.\ observations that are not in any cluster.  Mixtures with a
uniform term were also used by Fraley and Raftery (1996).  The
parameter $k$ controls the complexity of the density $f$.  Larger
values of $k$ allow us to approximate very complex densities $f$ but
also entail estimating many more parameters.

It is important to emphasize that we are not assuming that the true
density $f$ is exactly of the form (\ref{eq:mix}). It suffices that
$f$ can be approximated by such a distribution of Gaussians.  For
large enough $k$, nearly any density can be approximated by such a
distribution.  This point is made precise in Genovese and Wasserman
(1998) and Roeder and Wasserman (1997).

\subsection{Estimating $\theta_k$ By Maximum Likelihood}

Assume first that $k$ is fixed and known.  For simplicity, we write
$\theta_k$ simply as $\theta$ in this section.  The most common method
for estimating $\theta$ is the method of maximum likelihood.  First,
the likelihood function ${\cal L}(\theta)$ is defined to be the
probability of the observed data $(x_1, \ldots , x_n)$ as a function
of the unknown parameters:
$$
{\cal L}(\theta) =f(x_1, \ldots , x_n ; \theta)
=\prod_{i=1}^n f(x_i ; \theta).
$$
Note that the data are now fixed at their observed values so that
${\cal L}(\theta)$ is a function over the parameter space $\Theta$.
The value $\hat{\theta}$ that maximizes the likelihood ${\cal
L}(\theta)$ is called the ``maximum likelihood estimator.''  In
passing we remark that maximizing ${\cal L}(\theta)$ is equivalent to
maximizing the log-likelihood $\ell (\theta) = \log {\cal L}(\theta)$
and it is usually easier to maximize the latter.  Maximum likelihood
estimators are known to enjoy certain optimality properties.  Loosely
speaking, one can show, under fairly weak assumptions, that they are
the most precise estimators available.  For a rigorous explanation,
see, for example, van der Vaart (1998, chapter 8).

In the case of mixture models, there are a few problems with maximum
likelihood estimation.  First, there are singularities in the
likelihood, that is, there are points $\theta$ in the parameter space
$\Theta$ where ${\cal L}(\theta)=\infty$.  These points occur at the
boundary of the space.  When we refer to the maximum likelihood
estimate we mean the maximum over the interior of $\Theta$ which thus
excludes these singularities.  Another problem is that ${\cal
L}(\theta)$ has many local maxima which is a fundamental numerical 
problem common to many such fitting and optimization procedures.   

\subsection{The EM Algorithm}

The next questions is: how do we find the maximum likelihood estimate
$\hat{\theta}$?  (We are continuing for the moment with the assumption
that $k$ is fixed and known.)  The usual method for finding
$\hat{\theta}$ is the EM (expectation maximization) algorithm.  We can
regard each data point $X_i$ as arising from one of the $k$ components
of the mixture.  Let $G=(G_1, \ldots , G_k)$ where $G_i=j$ means that
$X_i$ came from the $j^{th}$ component of the mixture.  We do not
observe $G$ so $G$ is called a latent (or hidden) variable.  Let
$\ell_c$ be the log-likelihood if we knew the latent labels $G$.  The
function $\ell_c$ is called the complete-data log-likelihood.  The EM
algorithm proceeds as follows.  We begin with starting values
$\theta_0$. We compute $Q_0 = E( \ell_c; \theta_0)$, the expectation
of $\ell_c$, treating $G$ as random, with the parameters fixed at
$\theta_0$. Next we maximize $Q_0$ over $\theta$ to obtain a new
estimate $\theta_1$. We then compute $Q_1 = E(\ell_c ; \theta_1)$ and
continue this process. Thus we obtain a sequence of estimates
$\theta_0, \theta_1, \ldots$ which are known to converge to a local
maximum of the likelihood.

We can be more explicit about the algorithm that results from this
process.
More detail is contained in the appendix to this paper.  
Define
$\tau_{ij}$ to be the probability that $X_i$ came from group $j$ given
the current values of the parameters: for $j\neq 0$ this is given by
\begin{equation}\label{tau}
\tau_{ij} = Pr( X_i \ {\rm is\ from\ Group\ } j) =
\frac{ p_j \phi( X_i ; \mu_j, \Sigma_j) }
     { p_0U(X_i)+ \sum_t p_t \phi( X_i ; \mu_t, \Sigma_t) }.
\end{equation}
For $j=0$ this is
\begin{equation}
\tau_{ij} = Pr( X_i \ {\rm is\ from\ Group\ } j) =
\frac{ p_0 U(X_i)}
     { p_0U(X_i)+ \sum_t p_t \phi( X_i ; \mu_t, \Sigma_t) }.
\end{equation}
Next find
\begin{eqnarray}
\hat{p}_j &=&
\frac{1}{n} \sum_{i=1}^n \tau_{ij}\\
\hat{\mu}_j &=& \frac{1}{n \hat{p}_j}\sum_{i=1}^n \tau_{ij} X_i\\
\hat{\Sigma}_j &=& \frac{1}{n \hat{p}_j}
\sum_{i=1}^n \tau_{ij} ( X_i - \hat{\mu}_j)( X_i - \hat{\mu}_j)^T.
\end{eqnarray}
Then we again compute the $\tau_{ij}'s$ etc.  A formal derivation of
this procedure is given in the McLachlan and Basford (1987).

\subsection{Choosing the Number of Components in the Mixture $k$}

So far, we have assumed that $k$ is known.  In fact, $k$ must also be
estimated from the data.  Large values of $k$ lead to highly variable
estimates of the density.  Small values of $k$ lead to more stable but
highly biased estimates.  Trading off the bias and variance is
essential for a good estimate.

One approach to choosing $k$ is to sequentially test a series of
hypotheses of the form

$H_0:$ number of components is $k$ versus

$H_1:$ number of components is $k+1$ 

\noindent
and repeat this for various values of $k$.  The usual test for
comparing such hypotheses is called the ``likelihood ratio test''
which compares the value of the maximized log-likelihood under the two
hypotheses.  This approach is infeasible for large data sets where $k$
might be huge.  Also, this requires knowing the distribution of the
likelihood ratio statistic.  This distribution is not known.  Large
sample approximations are available (Dacunha-Castelle and Gassiat
1996) but these are unwieldy.

Here we summarize three promising approaches.  In what follows we let ${\cal
F}_k$ denote the set of all densities that are mixtures of $k$ normals.
Suppose that for each $k$ we have obtained the maximum likelihood estimator
$\hat{\theta}_k$ of $\theta_k$.  (In practice, we do not implement the method
this way but for now let us assume that we do.)  Let $\hat{f}_k(\cdot)=
f(\cdot \ ;\hat{\theta}_k)$.  Also, let $\ell_k(\hat{\theta}_k) = \sum_{i=1}^n
\log \hat{f}_k(X_i)$ be the value of the log-likelihood function at the
maximized parameter value.  Finally, we let $f_0$ denote the true density
function which generated the data.  We now have a set
$\{\hat{f}_1,...,\hat{f}_k, \ldots\}$ of possible density estimates available
to us and we want to choose one of them.  The first two methods choose $k$ by
maximizing quantities of the form $\ell(\hat{\theta}_k) - \lambda_k R_k$ where
$R_k$ is the number of parameters in the $k^{th}$ model; this is a penalised
log--likelihood of which there are many choices. For this paper, we have considered
the two most common penalized log--likelihoods, namely taking $\lambda_k=1$
gives the AIC criterion while taking $\lambda_k=\log n /2$ gives the BIC
criterion. 

\subsubsection{Akaike Information Criterion (AIC).}  
Suppose our goal is to choose $k$ so that $\hat{f}_k$ is close to
$f_0$.  In the fields of Statistics and
Information Theory, 
a common way to measure closeness is through the
Kullback-Leibler (KL) divergence defined by
$$
D(f,g) = \int f (x) \log \frac{ f(x)}{g(x)} dx =
\int f (x) \log  f(x) dx -\int f (x) \log g(x) dx.
$$
It can be shown that $D(f,g)\geq 0$ and $D(f,g)=0$ if $f=g$.  
Suppose then that we want to choose $k$ to
minimize $D(f_0,\hat{f}_k)$.  It is easy to see that this is
equivalent to choosing $k$ to maximize
$$
H(f_0, \hat{f}_k) \equiv \int f_0 (x) \log \hat{f}_k(x)
$$
since $D(f_0,\hat{f}_k) = c - H(f_0, \hat{f}_k)$ where $c = \int
f_0(x) \log f_0(x) dx$ is a constant which is common to all the models
being compared.  Note that $H(f_0, \hat{f}_k)$ is an unknown quantity
since it involves the true density $f_0$ which is the very thing we
are trying to estimate.  So we must estimate $H(f_0, \hat{f}_k)$ for
each $k$ and then choose $k$ to maximize the estimate.

To this end, define
$$
V_k = \ell(\hat{\theta}_k)-  R_k
$$
where $R_k$ is the number of parameters in the model.  This quantity
is called AIC (Akaike Information Criterion, Akaike 1973).  For
certain statistical models it can be shown that
\begin{equation}\label{eq:aic}
\frac{1}{n}V_k =H(f_0, \hat{f}_k)+ O_P\left(\frac{1}{\sqrt{n}}\right)
\end{equation}
which implies that maximizing $V_k$ corresponds approximately to
maximizing the desired quantity $H(f_0,\hat{f}_k)$.  Thus, we choose
$k$ maximizing $V_k$.  In the above formula, $Z_n = O_P(a_n)$ means
that, for every $\epsilon >0$, there exists $B$ such that 
$Pr( | Z_n/a_n | > B)< \epsilon$, for large $n$.

Unfortunately, the conditions needed to justify (\ref{eq:aic}) do not
hold in mixture models.  This does not mean that AIC will not work
but only that a mathematical proof is still not available.
Barron and Yang (1995) and Barron, Birg\'{e} and Massart (1999) give
some very general results about model selection that suggest that AIC
(or a modified version of AIC) should work well.  They show that if
$\lambda_k$ is chosen appropriately, and $\hat{k}$ is chosen to
maximize $\ell (\hat{\theta}_k) - \lambda_k R_k$, then
$$
E( d^2 (f_0,\hat{f}_{\hat{k}})) \leq c \inf_k \left\{ \inf_{f \in
{\cal F}_k} D(f_0,f) + \frac{\lambda_kR_k}{n}\right\}
$$
for some $c>0$, where $d^2(f,g) = \int ( \sqrt{f}(x) - \sqrt{g}(x))^2
dx$.  This means that the estimated density will be close to the true
density if an AIC-like quantity is used, even in cases where the
justification cited earlier does not apply.  The correct values for
$\lambda_k$ for mixture models have not yet been determined.  Some
preliminary results in this direction are available in Genovese and
Wasserman (1998).  However, the lack of firm theoretical results
emphasizes the need to study these methods by way of simulation.  For
now we use $\lambda_k =1$.

\subsubsection{Bayesian Information Criterion (BIC).}  
A competitor to AIC is BIC (Bayesian Information Criterion, Schwarz
1978).  Instead of maximizing $V_k$ we instead choose $k$ to maximize
$$
W_k = \ell(\hat{\theta}_k)- \frac{R_k}{2} \log n.
$$
Whereas AIC is designed to find a $k$ that makes $\hat{f}_k$ close to
$f_0$, BIC is designed to find the ``true value'' of $k$.  Let us
explain what we mean by the true value of $k$.

Note that the models are nested, i.e.\  ${\cal F}_1 \subset {\cal F}_2
\subset \cdots$.  Suppose that there is a smallest integer $k_0$ such
that $f_0\in {\cal F}_{k_0}$.  Let $\hat{k}_n$ denote the value of $k$
that maximizes $W_k$.  In many cases, it can be shown that BIC is
asymptotically consistent, meaning that
$$
Pr( {\rm there\ exists\ }n_0\ {\rm such\ that, \ } \hat{k}_n = k_0 \
{\rm for\ all\ }n\geq n_0) =1.
$$
In other words, if there is a true model (the smallest model
containing $f_0$), then BIC will eventually find it.  This consistency
property is known to hold in many statistical models.  It was only
recently shown to hold for mixture models (Keribin, 1999).  Despite
the appeal of consistency, we note some drawbacks.  First, there may
not be such a $k_0$.  It might be that $f_0$ is not contained in any
${\cal F}_k$. In this case it is still meaningful to use the models to
find a good approximation to $f_0$, but it is not meaningful to
estimate $k_0$.  Second, we note that finding $k_0$ (assuming it even
exists) is not the same as finding an estimate $\hat{f}_k$ which is
close to the true density $f_0$.  One can show that there are cases
where $D(f_0, \hat{f}_k) < D(f_0, \hat{f}_{k_0})$ for some $k\neq
k_0$.  In other words, finding a good density estimate and identifying
the true model are different goals.

We should also note that BIC has a Bayesian interpretation (hence its
name).  Specifically, maximizing $W_k$ corresponds approximately to
maximizing to the posterior probability of the $k^{th}$ model
$Pr({\cal F}_k| X^n)$.  Again, this fact has been proved for some
statistical models but not for mixture models.

\subsubsection{Data Splitting.}  
AIC is aimed at getting an approximately unbiased estimate of $H(f_0,
\hat{f}_k)=\int f_0(x) \hat{f}_k (x)dx$.  An alternative method for
estimating $H(f_0, \hat{f}_k)$ is to split the data into two parts
$X^n =(Y,Z)$ where $Y$ is of length $n_1$ and $Z$ is of length $n_2$
and $n=n_1 + n_2$.  The models are fit using $Y$ and then we estimate
$H(f_0, \hat{f}_k)$ with $Z$ by using
$$
T_k =\frac{1}{n_2} \sum_{i=1}^{n_2} \log \hat{f}_k (Z_i).
$$
It follows immediately that $T_k$ is an unbiased estimate of $H(f_0,
\hat{f}_k)$, i.e.\  $E(T_k) = H(f_0, \hat{f}_k)$.  We then choose $k$
to maximize $T_k$.

There is a tradeoff here. If $n_1$ is small, then the estimates
$\hat{f}_k$ are based on little data and will not be accurate. If
$n_2$ is small then $T_k$ will be a poor estimate of $H(f_0,
\hat{f}_k)$: it is unbiased but will have a large variance.  Smyth
(1996), building on work of Burman (1989) and Shao (1993), suggests
one possible implementation of this method.  Split the data randomly
into two equal parts and compute $T_k$ as above.  Now repeat this many
times, choosing the splits at random, then average the values of $T_k$
and choose $k$ from these averaged values.  This method shows some
promise but it is much more computationally intensive than the other
methods.

\subsection{Measurement Error.}  
Suppose we cannot observe $X_i$ but instead we observe $X_i^* = X_i +
\epsilon_i$ where $\epsilon_i \sim N(0, T_i)$.  For example, with
three dimensional spatial galaxy data, the third coordinate (the
red-shift) is measured with error.  The observed data are $(X_1^*,
\ldots , X_n^*)$.  This can be thought of as a two stage model:
\begin{eqnarray*}
X_i & \sim & {\rm Mixture\ of \ Gaussians}\\
X_i^* |X_i & \sim & N(X_i , T_i).
\end{eqnarray*}
The likelihood function is
$$
{\cal L}(\theta) =\prod_{i=1}^n g(x_i^* ; \theta)
$$
where
$$
g(x^*_i ; \theta) = \int \phi( x^*_i ; x , T_i) f(x ; \theta) dx.
$$
Here, $f(x ; \theta)$ is the density from (\ref{eq:mix}).  The maximum
likelihood estimator is defined as before as the value that maximizes
${\cal L}(\theta)$ and fortunately, it is easy to adapt the EM
algorithm for this new likelihood function.

Let us assume that $T_i =T$ is the same for each observation.  The
extension to the case where $T_i$ varies for each observation is
easily handled and will be discussed in a future paper.  The maximum
likelihood estimates can be found as follows: we run the EM algorithm.
Let $(p^*, \mu^*, \Sigma^*)$ denote the estimates that result after EM
has converged.  Then the maximum likelihood estimates $\hat{p},
\hat{\mu}$ and $\Sigma$ are given by $\hat{p} = p^*$, $\hat{\mu} =
\mu^*$ and
$$
\hat{\Sigma}_j =
\left\{
\begin{array}{ll}
{\Sigma}_j^* - T & {\rm if\ }\Sigma_j^* - T \ 
{\rm is\ positive\ definite }\\
Z & {\rm otherwise}
\end{array}
\right.
$$
where $Z$ is the matrix whose entries are all 0.

\subsection{Comparison to Kernel Density Estimators}

Another way to estimate the density $f$ is by way of the kernel
density estimate defined by
$$
\hat{f} (x) = \frac{1}{n} \sum_{i=1}^n
\frac{1}{h_n}K\left( \frac{x-X_i}{h_n}\right).
$$
(For simplicity, we are considering $x$ to be one-dimensional here.)
In the above equation, $K(t)$ is a kernel, that is, a function which
satisfies $K(t)\geq 0$, $\int K(t) dt =1$ and $\int tK(t) dt =0$.  A
common choice of kernel is the Gaussian kernel $K(t) = \{ 2
\pi\}^{-1/2} e^{-t^2/2}$.  The bandwidth $h_n$ controls the amount of
smoothing.  It is well known that the choice of kernel is unimportant
but that the choice of bandwidth is crucial.  There is a large
literature on choosing $h_n$ optimally from the observed data.

Kernel density estimates are like mixture models but the complexity is
controlled via the bandwidth instead of the number of components $k$.
Kernels have the advantage of being theoretically well-understood and
they can be computed quickly using the fast-Fourier transform.  On the
other hand, using a fixed bandwidth puts severe restrictions on the
density estimate.  In this sense, mixture models are more adaptive
since different covariance matrices are used in different regions of
the sky.  In principle, one can also make kernel density estimates
more adaptive by allowing the bandwidth to vary with each data point.
To date, adaptive bandwidth kernel density estimation has not been a
thriving practical success because of the difficulties in choosing
many bandwidths.  Indeed, mixture models may be regarded as a
practical alternative to adaptive bandwidth kernel density estimators.
There are other adaptive density estimators, such as wavelets, which
are beyond the scope of this paper.

\section{Computational Issues}

Given the definition of the EM algorithm and the criteria for
determining the number of components to the model using AIC and BIC we
must now address how do we apply this formalism to massive
multidimensional astronomical datasets. In its conventional
implementation, each iteration of EM visits every data point-class
pair, meaning $NR$ evaluations of a $M$-dimensional Gaussian, where
$R$ is the number of data-points and $N$ is the number of components
in the mixture. It thus needs $O(M^2 NR)$ arithmetic operations per
iteration. For data sets with 100s of attributes and millions of
records such a scaling will clearly limit the application of the EM
technique. We must,therefore, develop an algorithmic approach that
reduces the cost and number of operations required to construct the
mixture model.

\subsection{Multi-resolution KD-trees}

An {\mrkdtree} (Multi-resolution KD-tree), introduced in Deng \& Moore (1995),
and developed further in Moore, Schneider \& Deng (1997) and Moore (1999), is
a binary tree in which each node is associated with a subset of the data
points. The root node owns all the data points.  Each non-leaf-node has two
children, defined by a splitting dimension $\nodesplitdim$ and a splitting
value $\nodesplitval$. The two children divide their parent's data points
between them, with the left child owing those data points that are strictly
less than the splitting value in the splitting dimension, and the right child
owning the remainder of the parent's data points:
\begin{eqnarray}
\xsubi \in \nodeleft & \Leftrightarrow &
\xsubi[\nodesplitdim] < \nodesplitval \mbox{ and } 
\xsubi \in \node \\
\xsubi \in \noderight & \Leftrightarrow &
\xsubi[\nodesplitdim] \geq \nodesplitval \mbox{ and } 
\xsubi \in \node
\end{eqnarray}
The distinguishing feature of {\mrkdtrees} is that their nodes contain
the following:
\begin{itemize}

\item
$\nodenumpoints$: The number of points owned by $\node$ (equivalently,
the average density in $\node$).

\item
$\nodecentroid$:
The centroid of the points owned by $\node$ (equivalently, the
first moment of the density below $\node$).

\item
$\nodecov$:
The covariance of the points owned by $\node$ (equivalently, the
second moment of the density below $\node$).

\item
$\nodehrect$:
The bounding hyper-rectangle of the points below $\node$ 
(not strictly necessary, but can lead to greater efficiency).
\end{itemize}

We construct {\mrkdtrees} top-down, identifying the bounding box of
the current node, and splitting in the center of the widest dimension.
A node is declared to be a leaf, and is left unsplit, if the widest
dimension of its bounding box is $\leq$ some threshold, $\mbw$. If
$\mbw$ is zero, then all leaf nodes denote singleton or coincident
points, the tree has $O(R)$ nodes and so requires $O(M^2 R)$ memory,
and (with some care) the construction cost is $O(M^2 R + M R \log
R)$. In practice, we set $\mbw$ to $1\%$ of the range of the data point
components. The tree size and construction thus cost considerably less
than these bounds because in dense regions, tiny leaf nodes were able
to summarize dozens of data points. Note too that the cost of
tree-building is amortized---the tree must be built once, yet EM
performs many iterations, and the same tree can be re-used for
multiple restarts of EM, or runs with differing numbers of components.

To perform an iteration of EM with the {\mrkdtree}, we call the
function {\makestats} (described below) on the root of the
tree. $\makestats(\node,\modelt)$ outputs $3N$ values: $( \swone ,
\swtwo , \ldots \swN , \swxone , \ldots \swxN , \swxxone , \ldots
\swxxN )$ where
\begin{equation}
\swj = \sum_{\xsubiinnode} \wij \commaspace
\swxj = \sum_{\xsubiinnode} \wij \xsubi \commaspace
\swxxj = \sum_{\xsubiinnode} \wij \xsubi \xsubi^T 
\end{equation}
The results of $\makestats(\rootnode)$ provide sufficient statistics to
construct $\modelsup{t+1}$:
\begin{equation}
\pj \assign \swj / R \commaspace
\muj \assign \swxj / \swj \commaspace
\covj \assign ( \swxxj / \swj ) - \muj \muj^T
\end{equation}

If {\makestats} is called on a leaf node, we simply compute, for each
$j$,
\begin{equation}
\wbarj = P(\classj \mid \xbar, \modelt) =
P(\xbar \mid \classj, \modelt) P(\classj \mid \modelt) /
\sum_{k = 1}^N 
P(\xbar \mid \classk, \modelt) P(\classk \mid \modelt)
\end{equation}
where $\xbar = \nodecentroid$, and where all the items in the right
hand equation are easily computed.  We then return $\swj = \wbarj
\times \nodenumpoints$, $\swxj = \wbarj \times \nodenumpoints \times
\xbar$ and $\swxxj = \wbarj \times \nodenumpoints \times \nodecov$.
The reason we can do this is that, if the leaf node is very small,
there will be little variation in $\wij$ for the points owned by the
node and so, for example $\sum \wij \xsubi \approx \wbarj \sum
\xsubi$.  In the experiments below we use very tiny leaf nodes,
ensuring accuracy.

If {\makestats} is called on a non-leaf-node, it can easily compute
its answer by recursively calling {\makestats} on its two children and
then returning the sum of the two sets of answers. In general, that is
exactly how we will proceed. If that was the end of the story, we
would have little computational improvement over conventional EM,
because one pass would fully traverse the tree, which contains $O(R)$
nodes, doing $O(NM^2)$ work per node. We will win if we ever spot
that, at some intermediate node, we can {\em prune}, i.e.\ evaluate the
node as if it were a leaf, without searching its descendents, but
without introducing significant error into the computation. To do this
we compute, for each $j$, the minimum and maximum $\wij$ that any
point inside the node could have. This procedure is more complex than
in the case of locally weighted regression (see Moore, Schneider \& Deng 1997).

\begin{figure}[htb]
\centerline{\psfig{file=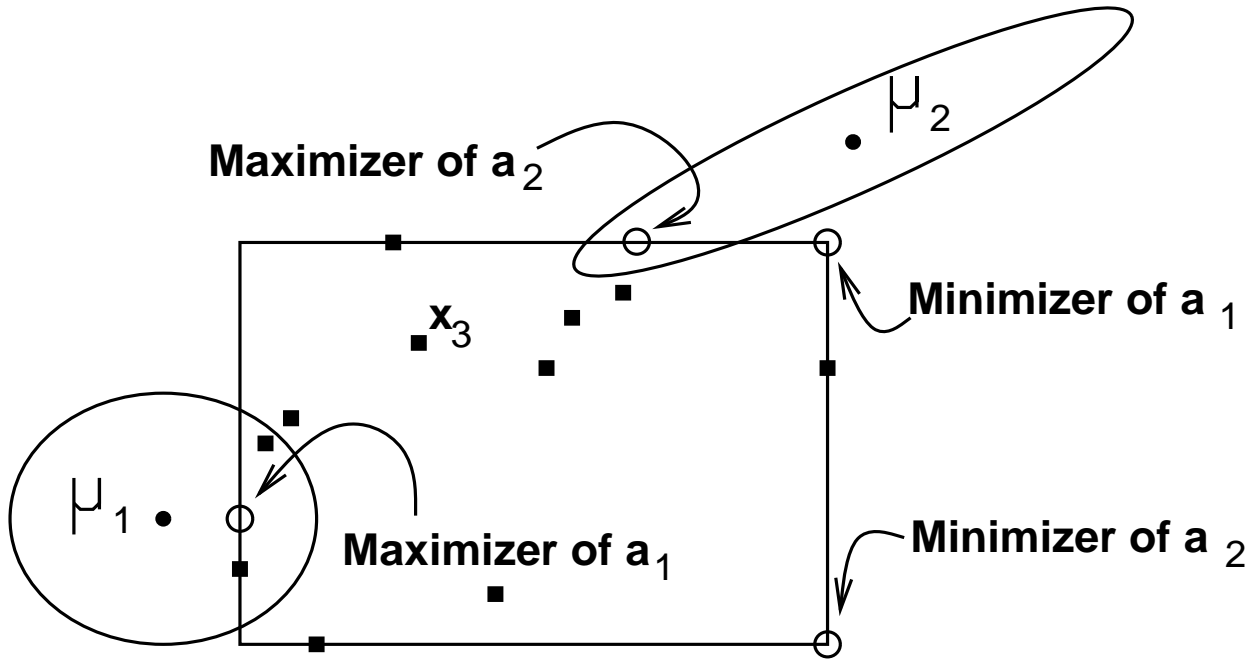,width=\textwidth}}
\caption{The rectangle denotes a hyper-rectangle in the
mrkd-tree. The small squares denote data points ``owned'' by the
node. Suppose there are just two classes, with the given means, and
covariances depicted by the ellipses.  Small circles indicate the
locations within the node for which $a_j$ (i.e.\ $P(x\mid\classj)$)
would be extremized.
\label{minmax}
}
\end{figure}

We wish to compute $\wjmin$ and $\wjmax$ for each $j$, where $\wjmin$
is a lower bound on $q\min_{\xsubiinnode} \wij$ and $\wjmax$ is an
upper bound on $\max_{\xsubiinnode} \wij$.  This is hard because
$\wjmin$ is determined not only by the mean and covariance of the
$j$th class but also the other classes. For example, in
Figure~\ref{minmax}, $\wthreetwo$ is approximately $0.5$, but it would
be much larger if $\classone$ were further to the left, or had a
thinner covariance.

The $\wij$'s are defined in terms of $\aij$'s, thus: $\wij = \aij \pj
/ \sum_{k = 1}^N \aik \pk$. We {\em can} put bounds on the $\aij$'s
relatively easily. It simply requires that for each $j$ we
compute\footnote{Computing these points requires non-trivial
computational geometry because the covariance matrices are not
necessarily axis-aligned. There is no space here for details.} the
closest and furthest point from $\muj$ within $\nodehrect$, using the
Mahalanobis distance $\mhdist(\boldx,\boldx') = (\boldx - \boldx')^T
\covj^{-1} (\boldx - \boldx')$.  Call these shortest and furthest
squared distances $\mhdmin$ and $\mhdmax$. Then
\begin{equation}
\ajmin = 
\gaussfactor \exp( -\half \mhdmax )
\end{equation}
is a lower bound for $\min_{\xsubiinnode} \aij$, with a similar
definition of $\ajmax$.
Then write 
\begin{eqnarray*}
\min_{\xsubiinnode} \wij & = & \min_{\xsubiinnode} (\aij \pj/ \sum_k \aik \pk)
= \min_{\xsubiinnode} (\aij \pj / (\aij \pj + \sum_{k \neq j} \aik \pk)) \\
& \geq & \ajmin \pj / ( \ajmin \pj + \sum_{k \neq j} \akmax \pk ) = \wjmin
\end{eqnarray*}
where $\wjmin$ is our lower bound. There is a similar definition for
$\wjmax$. The inequality is proved by elementary algebra, and requires
that all quantities are positive (which they are).  We can often
tighten the bounds further using a procedure that exploits the fact
that $\sum_j \wij = 1$, but space does not permit further discussion.

We will prune if $\wjmin$ and $\wjmax$ are close for all $j$.  What
should be the criterion for closeness? The first idea that springs to
mind is: Prune if $\forall j \ . \ (\wjmax - \wjmin < \epsilon)$.  But
such a simple criterion is not suitable: some classes may be
accumulating very large sums of weights, whilst others may be
accumulating very small sums. The large-sum-weight classes can
tolerate far looser bounds than the small-sum-weight classes. Here,
then, is a more satisfactory pruning criterion: Prune if $\forall j \
. \ (\wjmax - \wjmin < \tau \wjtotal)$ where $\wjtotal$ is the total
weight awarded to class $j$ over the entire dataset, and $\tau$ is
some small constant. Sadly, $\wjtotal$ is not known in advance, but
happily we can find a lower bound on $\wjtotal$ of $\wjsofar +
\nodenumpoints \times \wjmin$, where $\wjsofar$ is the total weight
awarded to class $j$ so far during the search over the {\kdtree}.

The algorithm as described so far performs
divide-and-conquer-with-cutoffs on the set of data points. In addition,
it is possible to achieve an extra acceleration by means of divide and
conquer on the class centers. Suppose there were $N = 100$
classes. Instead of considering all 100 classes at all nodes, it is
frequently possible to determine at some node that the maximum
possible weight $\wjmax$ for some class $j$ is less than a miniscule
fraction of the minimum possible weight $\wkmin$ for some other class
$k$. Thus if we ever find that in some node $\wjmax < \lambda \wkmin$
where $\lambda = {10}^{-4}$, then class $\classj$ is removed from
consideration from all descendents of the current node.  Frequently
this means that near the tree's leaves, only a tiny fraction of the
classes compete for ownership of the data points, and this leads to
large time savings.

\subsection{A Heuristic Splitting Approach to Choosing the Number of
Components.}

We possess the mathematical and computational tools needed to implement a
Mixture Model. The final tasks is to determine $k$, the number of components
in the model.  The conventional solution to this problem is to run EM with one
Gaussian, then two Gaussians, then three up to some maximum and choose the
resulting mixture that minimizes our scoring metric (for example BIC, AIC or a
test set log-likelihood).  

\begin{figure}[htb]
\centerline{\psfig{file=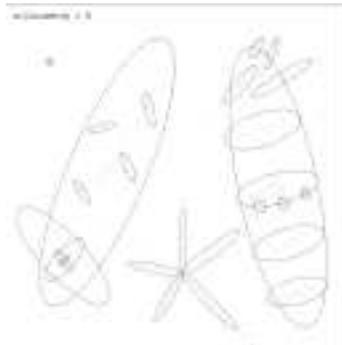}}
\caption{two-dimensional Gaussian mixture. The ellipses show the
two-sigma contours of the ellipses. The dots show a sample
of the data from the mixture (the full dataset has 80000 points).
\label{connect-create}
}
\end{figure}

Consider the distribution and Gaussians shown in
Figure~\ref{connect-create}.  What happens when we try to
estimate the underlying distribution from data?  The true number of
Gaussians that created this data is 27, but the EM algorithm does not
know this in advance (and even if it did, there is no guarantee that
running EM starting with 27 Gaussians will minimize the expected
divergence between the true and estimated distribution).
We show the results of doing so in Table 1.

There are a number of computational and optimization reasons for not
simply increasing the number of Gaussian components until the the AIC
or BIC score is minimized. Computationally choosing the best $k$ out
of a range $1 \ldots k_{\myindex{max}}$ is expensive as it costs
$k_{\myindex{max}}$ times the computation of a single EM run.  In
terms of optimization even for the $k$ that would in principal be able
to achieve the best score, the local-optima problem means we are
unlikely to find the best configuration of Gaussians for that $k$.
This can easily be seen in Figures~\ref{bic-seq-gauss}
and~\ref{aic-seq-gauss} where some of the Gaussians are
clearly needed in another part of space, but have no means to ``push
past'' other Gaussians and get to their preferred locations.

To combat these problems, we use an elementary heuristic splitting
procedure. 
(A similar algorithm, for Bayesian computations is contained in
Richardson and Green, 1997.)
We begin by running with one Gaussian. We then execute the
following algorithm:

\begin{verbatim}
1. Randomly decide whether to try increasing or decreasing the number
   of Gaussians.

2. If we decide to increase...

   2.1 Randomly choose a number between 0 and 1 called SPLITFRACTION

   2.2 Some of the Gaussians will be allowed to split in two. We sort
       to Gaussians in increasing order of a quantity known as SPLITCRITERION.

   2.3 The best N Gaussians according to this criterion are allowed
       to split, where N = SPLITFRACTION * Total Number of Gaussians

   2.4 Each splitting Gaussian is divided into two, by making two copies
       of the original, squashing the covariance to become thinner in the
       direction of the principal component of the original covariance
       matrix, and then pushing one copy's mean a short distance in one
       direction along this component and the other copy's mean in the other
       direction. Figure 1 shows an example.

3. If we decide to decrease...

   3.1 Randomly choose a number between 0 and 1 called KILLFRACTION

   3.2 Some of the Gaussians will be deleted. We sort
       to Gaussians in increasing order of a quantity known as KILLCRITERION.

   3.3 The worst N Gaussians according to this criterion are deleted
       where N = KILLFRACTION * Total Number of Gaussians

4. Then, whether or not we decided to increase or decrease, we run EM
   to convergence from the new mixture. We then look at the model scoring
   criterion (AIC, BIC or testset) for the resulting mixture. Is it better
   than the original mixture?

5. If it's better...

   5.1 We restart a new iteration with the new mixture and jump to Step 1.

6. If it's worse...

   6.1 We revert to the mixture we had at the start of the iteration and
       jump to Step 1.
\end{verbatim}

In the following experiments, SPLITCRITERION and KILLCRITERION are
very simple: we split Gaussians with relatively high mixture
probabilities and we kill Gaussians with relatively low mixture
probabilities. We have performed experiments (not reported here) with
alternative, and occasionally exotic, criteria, but find that the
simple approach reported here is generally reasonable.

This approach benefits from searching through many values of $k$ in large
steps, and so is faster, But more importantly it removes some of the
local minima problems, so that a suboptimal local optimum with $k$ centers
may manage to move to a superior solution with $k$ centers by means
of an intermediate step at which useful new centers are added, iterated,
and then less useful centers are deleted.

\section{Applications to Astrophysical problems}

We consider here two astrophysical applications of the density
estimation techniques presented in this paper to illustrate their
potential (though as we note below there are a large number of
applications that would benefit from this approach). In the next
section we apply mixture models to the reconstruction of the density
distribution of galaxies from photometric and spectroscopic
surveys. In the subsequent sections we discuss the application 
of mixtures to
modeling the distribution of stellar colors in order to identify
objects with anomalous colors (e.g.\ searching for high redshift
QSOs).

\subsection{Reconstructing the Density Distribution of Galaxies}
\label{reconstruct}

Observations of the distributions of galaxies have shown that the dominant
large-scale features in the Universe are comprised of walls and voids (Geller
and Huchra 1990, Broadhurst et al. 1990). These structures exist over a wide
range of scales, from filaments that are a few \Mpc\ across to walls that
extend across the size of the largest surveys we have undertaken. The
consequence of this is that any technique that is used to measure the density
distribution must be able to adapt to this range of scales.

In this section we compare the EM algorithm, as presented in this
paper, to a more traditional kernel--smoothing density estimator. For
this comparison, we use simulated data sets generated from a Voronoi
Tessellation since the observed distribution of filaments and sheets of
galaxies in the universe can be reasonably well simulated using such a
distribution.  In this scenario the walls of the foam are considered
as the positions of the filaments and the underdense regions between
these walls the position of the voids. This simulation provides a very
natural description of structure formation in the Universe as a
Voronoi distribution arises when mass expands away from a set of
points until it collides with another overdensity and ceases
expanding.

In Figure 8, we present the underlying Voronoi density map we have
constructed; this is the ``truth'' in our simulation (the top--left
panel). From this distribution we derive a set of 100,000 data points to
represent a mock 2-dimensional galaxy catalog (top--right panel). We note here
we have not added any additional noise to this mock galaxy catalog so it is an
over--simplification of a real galaxy survey which would have
``fingers--of--God'' and measurement errors.

We have applied the EM algorithm (with both the AIC and BIC splitting
criteria) and a standard fixed kernel density estimator to these point-like
data sets in order to reproduce the original density field.  The latter
involved finely binning the data and smoothing the subsequent grid with a
binned Gaussian filter of fixed bandwidth which was chosen by hand to minimize
the KL divergence between the resulting smoothed map and the true underlying
density distribution ({\it i.e.\ } to minimize the difference between the
top--left and the bottom-left panels in Figure 8.  Clearly, we have taken the
optimal situation for the fixed kernel estimator since we have selected it's
bandwidth to ensure as close a representation of the true density distribution
as possible. This would not be the case in reality since we would never known
the underlying distribution and would thus be forced to choose the bandwidth
using various heuristics.

The lower left panel of Figure \ref{compare} shows the reconstructed density
field using the fixed kernel and the lower right panel the corresponding
density field derived from the EM algorithm using the AIC criterion.  As we
would expect the fixed kernel technique provides an accurate representation of
the overall density field. The kernel density map suffers, however, when we
consider features that are thinner than the width of the kernel. Such
filamentary structures are over-smoothed and have their significance
reduced. In contrast the EM algorithm attempts to adapt to the size of the
structures it is trying to reconstruct. The lower right panel shows that where
narrow filamentary structures exist the algorithm places a large number of
small Gaussians. For extended, low frequency, components larger Gaussians are
applied. The overall result is that the high frequency components (e.g.\ sharp
edges) present within the data are reproduced accurately with out the
smoothing of the data seen in the fixed kernel example.

To quantify these statements we use the KL divergence to test the similarity
between the different maps. For the fixed kernel estimator, we measure a KL
divergence of 0.074 between the final smoothed map and the true underlying
density map (remember, this is the smallest KL measurement by design).  For
the EM AIC density map we measure a KL divergence of 0.067 which is lower than
the best fixed kernel KL score thus immediately illustrating the power of the
EM methodology. We have not afforded the same prior knowledge to the EM
measurement -- {\it i.e.\ } hand--tune it so as to minimize the KL divergence
-- yet we have beaten the kernel estimator. For reference, the EM BIC
estimator scored 0.078 and a uniform distribution scored 2.862 for the KL
divergence with the true underlying density distribution.

In Figure \ref{compare1}, we compare the relative performances of the
EM AIC (top two panels) and BIC (lower two panels) criteria. The left
hand side of the figure shows the centers, sizes and orientations of
the Gaussians that are used to reconstruct the density distribution.
The right hand panel shows the reconstructed density field derived
from these Gaussians. The AIC criteria under smooths the data using a
larger number of small Gaussians to reproduce the observed density
field. While this better traces the small narrow features within the
data it is at the cost of a larger number of components (and
consequently a loss in the compactness of the representation of the
data). BIC in contrast converges to a smaller number of Gaussians and
a more compact representation. The fewer components (typically larger
Gaussians) do not trace the fine structure in the data.

\subsubsection{Application to the Las Campanas Redshift Survey}
Extending the analysis in Section \ref{reconstruct} to the case of the spatial
distribution of galaxies, we consider the distribution of galaxies taken from
the Las Campanas Redshift Survey (LCRS; Shectman et al. 1996). The LCRS
comprises of a spectroscopic and photometric survey of over 26,000 galaxies
selected from six regions of the sky. Each region comprises of a slice on the
sky approximately 1.5 degrees thick and 80 degrees wide. With a mean redshift
of $z=0.1$, the depth of this survey is approximately 300
\Mpc. Details of the selection function for the LCRS, the distribution of
galaxies on the sky and how this survey was constructed can be found in
Shectman et al. (1996) and Lin et al. (1996).

For the purpose of our analysis, we select a subset of the LCRS data which is
the co--addition of the data in the three ``northern'' slices of the survey
(slices at $-3^{\circ}$, $-6^{\circ}$ \& $-12^{\circ}$ in
declination). Moreover, we have ignored the thickness of the slice and
converted the original spherical coordinates to a two--dimensional Euclidean
coordinate system. We limited the analysis to a box $100\times150$\Mpc\ in
size taken from this data which contained 11195 galaxies in total; the
distribution of these galaxies is shown as (red) data points in Figure
\ref{LCRS} along with our EM analysis of these data.

The density maps (grayscales) shown in Figure \ref{LCRS} were obtained by
running the EM algorithm for 2 hours on a 600MHz PentiumIII machine with 128
MBytes of memory.  In Figure \ref{time}, we show the AIC score found by the EM
algorithm as a function of run time. The algorithm asymptotically approaches
the best AIC score while checking for local minima along the way (the ends of
the error bars in Figure \ref{time} show the starting and end AIC score after
one iteration of splitting procedure). For this case, we have used a splitting
probability of 0.125 and fixed the number of sub-iterations to be five, {\it
i.e.\ } the algorithms continues for 5 iterations along a given branch before
determining if this iteration has worsened the overall AIC score.  If so, the
EM algorithm returns to the original Gaussian centroids and tries again. This
behavior is clearly seen in \ref{time} where in some cases the ending AIC (top
of the error bar) is substantially worse than the beginning AIC
score. Therefore, this heuristic splitting algorithm ensures we approach the
true minimum AIC score in less than approximately 20 minutes for over 10,000
data points.
 
Figure \ref{LCRS} demonstrates the adaptive nature of EM. Here we
discuss practical issues of applying the EM algorithm to the LCRS
data; we present results of three different runs of the EM algorithm
with different initial conditions {\it i.e.\ } AIC, BIC and AIC plus a
uniform background. We do not present the combination of BIC and a
uniform background since in all cases the EM algorithm quickly
iterated to zero uniform background for these data.  In the case of
AIC scoring with no uniform background, the EM algorithm provides an
accurate representation of the data. Our only concern is that the
algorithm has potentially under-smoothed the data and does exhibit a
fraction of highly elongated Gaussians. These issues we explored in
section \ref{reconstruct} where we demonstrated that EM AIC was better
than even than the best kernel density estimator and also possessed
highly elongated Gaussians. These problems are lessened in the case of
AIC plus a uniform background but in this case, it does not accurately
represent the low density areas of the data which are probably better
described using a higher order functional form for the background
rather than the uniform component assumed here. In the case of BIC, EM
appears to have over-smoothed the data. It gives an accurate
representation of the large--scale structure in the data, but no
information on the small-scale structure.

In summary, the combination of AIC, BIC and some large--scale slowly
varying smooth background should provide the best representation of
the LCRS data.  This should be possible to implement since: {\it a)}
we can modify the penalized likelihoods as we wish -- Abramovich et
al. (2000) have recently proposed $\lambda_k=\log (\frac{R_k}{n})$ as
a new penalized likelihood that appears to live between the two
extremes of BIC and AIC (this will be explored in a future paper);
{\it b)} we can add more terms to the uniform background component of
the mixture model.  Specifically, we could initially compress the data
into a single redshift histogram and fit a low--order polynomial to
this distribution and use that as the initial background for the EM
algorithm. Alternatively, the large scale shape of the redshift
histogram is governed by selection effects; at low redshift it is
simply the volume increasing (as $z^2$) while at large redshifts it is
due to the magnitude limit which cuts--off the luminosity
function. These effects could be modelled and used as the initial
background. 

In Figure \ref{threshold}, we present a thresholded EM density map for
both AIC and BIC.  The threshold corresponds to a $3\sigma$
probability limit for these mixture models of Gaussian {\it i.e.\ } a
probability of greater than 0.001.  This shows the ability of EM to
accurately characterize highly non--linear hierarchical structures by
inverting the point source data to obtain the underlying density
distribution. This figure again illustrates the advantage of merging
these two representations of the density field since the BIC gives an
excellent representation of the large--scale structure, while the AIC
encapsulates the higher--order structure.

\subsection{Classification of Multicolor Data}

The EM algorithm is more generally applicable to density estimation
problems in astrophysics than the standard spatial clustering analyses
that we have discussed previously. We demonstrate this by applying EM
to the question of how to identify anomalous records in
multidimensional color data (e.g.\ the identification of high redshift
QSOs from color-color diagrams). In Figure~\ref{color} we show the
distribution of 6298 stellar sources in the B-V and V-R color space
(with R$<$22) taken from a 1 sq degree multicolor photometric survey
from Szokoly et al (2000). Applying the EM algorithm using the AIC
scoring criteria we derive the density distribution for the B-V -- V-R
color space. The grayscale image on Figure~\ref{color} shows the
resultant mixture density map. As we would expect from the results from our
analysis of the spatial distribution of galaxies in the LCRS the EM
distribution traces the stellar locus with the most dense regions on
this map reflecting the distribution of M stars.

>From the mixture density distribution we define a probability density
map. This gives the probability that a stellar object drawn at random
from the observed distribution of stars would have a particular set of
B-V -- V-R colors. We can now assign a probability to each star in the
original data that describes the likelihood that it arises from the
overall distribution of the stellar locus. We can then rank order all
sources based on the likelihood that they were drawn from the parent
distribution. The right panel of Figure~\ref{color} shows the colors
of the 5\% of sources with the lowest probabilities. These sources lie
preferentially away from the stellar locus. As we increase the cut in
probability the colors of the selected sources move progressively
closer to the stellar locus.

The advantage of the mixture approach over standard color selection
techniques is that we identify objects based on the probability that
they lie away from the stellar locus (i.e.\ we do not need to make
orthogonal cuts in color space as the probability contours will trace
accurately the true distribution of colors). While for two dimensions
this is a relatively trivial statement as it is straightforward to
identify regions in color--color space that lie away from the stellar
locus (without being restricted to orthogonal cuts in color-color
space) this is not the case when we move to higher dimensional data.
For four and more colors we lose the ability to visualize the data
with out projecting it down on to a lower dimensionality subspace
(i.e.\ we can only display easily 3 dimensional data). In practice we
are, therefore, limited to defining cuts in these subspaces which may
not map to the true multidimensional nature of the data.  The EM
algorithm does not suffer from these disadvantages as a probability
density distribution can be defined in an arbitrary number of
dimensions. It, therefore, provides a more natural description of both
the general distribution of the data and for the identification of
outlier points from high dimensionality data sets. With the new
generation of multi-frequency surveys we expect that the need for
algorithms that scale to a large number of dimensions will become more
apparent.

\section{Discussion \& Conclusions}

With the current and future generation of wide-angle, multi-frequency
surveys it is becoming increasingly apparent that we need to develop
new analysis techniques that scale to the domain of large numbers of
objects and multiple dimensions (e.g.\ colors). In this paper we have
presented one such statistical approach; the use of Mixture Models in
the density estimation of large astronomical datasets. Implementing
Mixture Models under a multi--resolution KD--tree framework we have
developed a fast, adaptive density estimator that can accurately
recover the underlying density distribution from point processes.

Applying these techniques to simulated galaxy redshift catalogs we
find that the EM algorithm provides a better representation of the
underlying density distributions than the best-case fixed-kernel
density estimators. We find that different scoring criteria for
determining the accuracy of the density estimation provide different
final representations of the underlying density. The BIC method
produces a smooth, compact representation of the data while AIC better
reflects high frequency components within the data with the associated
loss in compactness (i.e.\ an increased number of Gaussians).

We demonstrate the broad class of astrophysical problems that would
benefit from such approach by considering two distinct cases; the
clustering of galaxies in redshift space using the LCRS data set and
the clustering of stars in color space to identify objects of unusual
color. We find that the adaptive nature of the EM algorithm enables an
accurate description of both the low and high frequency components
within the LCRS data. As with the simulated data sets using the BIC
scoring results in an apparent under-smoothing of the data while the
AIC score produces a better representation of the high frequency modes
at the cost of requiring a larger number of Gaussians to describe the
data. A hybrid scoring criteria that mixes the properties of AIC and
BIC may provide the most accurate representation.

For applications to clustering in color space the results are equally
encouraging. As the result of the EM algorithm is a density map that
can be expressed in terms of the probability than an object would have
a particular set of colors EM provides a simple and intuitive approach
to identifying outliers within this color space. It can be implemented
for data of an arbitrary number of dimensions (and parameters) and is
therefore well suited to observations taken over a wide spectral
range.

To date we have only considered here a mixture model of Gaussians.  It is
mathematically straightforward to consider a mixture of any profiles one
desires. The price one may pay in that case is in efficient computation.  Two
astronomically interesting profiles we may try are the Plummer or isothermal
profile for identifying clusters of galaxies -- which has been used in
automated searches for clusters of galaxies -- and the double Gaussian profile
which is typically used to create a Gaussian--like profile but with long
wings.  The latter has been used to describe the point--spread function of the
Sloan Digital Sky Survey as well as other telescopes and instruments. We plan
to explore these other mixture models in future papers.

To facilitate the use of EM in astronomical data-analysis, a version of the
software used herein (FASTMIX) can be obtained from {\em
http://www.cs.cmu.edu/\~AUTON/applications.html}. At present, only a Linux
binary executable of the software has been made available but encourage people
to contact us directly if they wish to examine the source code or require a
binary built for a different operating system (email: nichol@cmu.edu).

\acknowledgments

We thank the National Science Foundation for supporting this research via
our multi--disciplinary KDI proposal ``New Algorithms, Architectures 
and Science for Data Mining of Massive Astrophysics Sky''. 

\appendix
\section{Derivation of the EM algorithm}

For the readers convenience, we include
here a derivation of the EM algorithm for the mixture model.
For simplicity of presentation, we omit the uniform component in what
follows.  The idea is to define a random variables $Z$ such that $f(x;
\theta) = \int f(x, z ; \theta)dz$ and such that maximizing the
complete data likelihood based on the $X_i's$ and the $Z_i's$ is easy.
We define $Q$ to be the expected value (over $Z$) of the complete data
log-likelihood.  We iterate between computing $Q$ and maximizing $Q$
For details, see Dempster, Laird and Rubin (1977).  Here are the
details for mixtures.

Define $Z_{ij} =1$ if $Y_i$ is from group $j$ and $Z_{ij}=0$
otherwise.  The log-likelihood of $Y_i's$ and the $Z_{ij}'s$ is called
the complete log-likelihood and is given by $\ell = \sum_j \sum_i
Z_{ij} [\log p_j + \log \phi(Y_i; \mu_j, \Sigma_j)]$.  Let
$\tilde{\theta}$ represent the current guess at the parameters.
Define
\begin{eqnarray*}
Q &=& E(\ell | Y^n , \tilde{\theta})\\ &=& \sum_j \sum_i E(Z_{ij} |
Y^n , \tilde{\theta}) [\log \tilde{p}_j + \log \phi(Y_i;
\tilde{\mu}_j, \tilde{\Sigma}_j)]\\ &=& \sum_j \sum_i \tau_{ij} [\log
\tilde{p}_j + \log \phi(Y_i; \tilde{\mu}_j, \tilde{\Sigma}_j)]
\end{eqnarray*}
where,
\begin{eqnarray*}
\tau_{ij} &=& E(Z_{ij} | Y^n , \tilde{\theta})\\ &=& Pr(Y_j \ {\rm is\
in\ group \ j} | Y^n , \tilde{\theta})\\ &=& \frac{f(y_i| {\rm group\
j })Pr( {\rm group\ }j)} {\sum_r f(y_i| {\rm group\ r })Pr( {\rm
group\ }r)}\\ &=& \frac{\tilde{p}_j \phi(y_i ; \tilde{\mu}_j,
\tilde{\Sigma}_j)} {\sum_r \tilde{p}_r \phi(y_i ; \tilde{\mu}_r,
\tilde{\Sigma}_r)}
\end{eqnarray*}
by Bayes' theorem.

Now, using the definition of a Gaussian, we have
\begin{eqnarray*}
Q &=& \sum_j \sum_i \tau_{ij} [\log p_j + \log \phi(Y_i; \mu_j,
\Sigma_j)]\\ &=& \sum_j \log p_j \sum_i \tau_{ij} - \frac{1}{2}\sum_j
\sum_i \tau_{ij} \log det \Sigma_j - \frac{1}{2}\sum_j\sum_i
\tau_{ij}(Y_i - \mu_j)^T \Sigma_j^{-1} (Y_i - \mu_j)\\ &=& \sum_j
\tau_{\cdot j} \log p_j - \frac{1}{2}\sum_j \tau_{\cdot j} \log det
\Sigma_j - \frac{1}{2}\sum_j\sum_i \tau_{ij}(Y_i - \mu_j)^T
\Sigma_j^{-1} (Y_i - \mu_j).
\end{eqnarray*}
where $\tau_{\cdot j} = \sum_i \tau_{ij}$. Let
$$
\hat{\mu}_j = \frac{ \sum_{i=1}^n \tau_{ij} Y_i}{\tau_{\cdot j}}.
$$
Note that $\sum_j \sum_i \tau_{ij} (Y_i - \hat{\mu}_j)^T \Sigma_j^{-1}
(\hat{\mu}_j -\mu_j) = 0$.  Also, let $tr A$ denote the trace of the
matrix $A$.  Recall that if $A$ is symmetric then $x^TA x = tr( A x^T
x)$.  Also, $tr (\sum_i A_i ) = \sum_i tr(A_i)$.  Therefore,
\begin{eqnarray*}
\sum_j\sum_i \tau_{ij}(Y_i - \mu_j)^T \Sigma_j^{-1} (Y_i - \mu_j)&=&
\sum_j \sum_i
( [Y_i - \hat{\mu}_j] + [\hat{\mu}_j -\mu_j])^T\Sigma_j^{-1} 
( [Y_i - \hat{\mu}_j] + [\hat{\mu}_j -\mu_j])\\
&=&
\sum_j\sum_i \tau_{ij}(Y_i - \hat{\mu}_j)^T \Sigma_j^{-1} (Y_i - \hat{\mu}_j)+
\sum_j \tau_{\cdot j} (\hat{\mu}_j - \mu_j)^T \Sigma_j^{-1} 
(\hat{\mu}_j - \mu_j)\\
&=&
\sum_j\sum_i 
tr \left\{\tau_{ij}\Sigma_j^{-1} (Y_i - \hat{\mu}_j)  (Y_i - \hat{\mu}_j)^T
\right\}+
\sum_j \tau_{\cdot j} (\hat{\mu}_j - \mu_j)^T \Sigma_j^{-1} 
(\hat{\mu}_j - \mu_j)\\
&=&
\sum_j tr \left(\Sigma_j^{-1} B_j\right) +
\sum_j \tau_{\cdot j} (\hat{\mu}_j - \mu_j)^T \Sigma_j^{-1} 
(\hat{\mu}_j - \mu_j)
\end{eqnarray*}
where
$$
B_j = \sum_i \tau_{ij} (Y_i - \hat{\mu}_j)  (Y_i - \hat{\mu}_j)^T.
$$
Hence,
$$
Q = \sum_j  \tau_{\cdot j} \log p_j -
\frac{1}{2}\sum_j \tau_{\cdot j} \log det \Sigma_j  -
\frac{1}{2} 
\sum_j tr \left( \Sigma_j^{-1} B_j\right) -
\frac{1}{2}
\sum_j \tau_{\cdot j} (\hat{\mu}_j - \mu_j)^T \Sigma_j^{-1} 
(\hat{\mu}_j - \mu_j).
$$

Now we need to maximize $Q$ subject to $\sum_j p_j =1$.  Take the
derivative with respect to $p_j$ and set equal to 0 to conclude that
the root satisfies
$$
\hat{p}_j =
\frac{\sum_i \tau_{ij}}{n}.
$$
Note that $\mu_j$ appears only in the last term.
This term is non-positive and is clearly maximized by setting
$\mu_j = \hat{\mu}_j$.
It remains then to find $\Sigma_j$ to maximize
$$
-\frac{1}{2}\sum_j  \tau_{\cdot j} \log det \Sigma_j  -
\frac{1}{2} 
\sum_j tr \left( \Sigma_j^{-1}B_j \right).
$$
Taking exponentials, note that this is the same as maximizing
\begin{equation}\label{ouch}
\sum_j \frac{1}{ (det \Sigma_j)^{\tau_{\cdot j}/2}}
\exp\left\{ - \frac{1}{2} tr \left(\Sigma_j^{-1} B_j \right)\right\}.
\end{equation}
In general, 
if $B$ is a $d \times d$ symmetric, positive definite matrix
and $b>0$ then
$$
\frac{1}{ (det (\Sigma))^b}
\exp\left\{ - \frac{1}{2} tr ( \Sigma^{-1} B)\right\} \leq
\frac{1}{ (det (B))^b}
(2b)^{db} e^{-bd}
$$
with equality if and only if $\Sigma = B/(2b)$.
It follows that
(\ref{ouch}) is maximized by
$$
\hat{\Sigma}_j =
\frac{B_j}{\tau_{\cdot j}}.
$$

\input{connolly.tab1.tex}

\clearpage

\begin{figure}[htb]
\centerline{\psfig{file=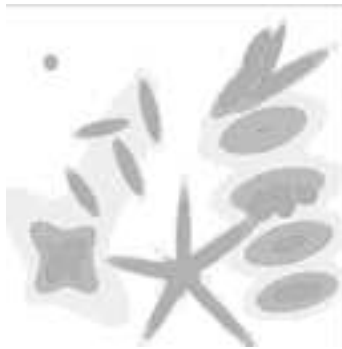}}
\caption{The
PDF of the mixture model of Figure~\ref{connect-create}
\label{connect-dmap}
}
\end{figure}

\begin{figure}[htb]
\centerline{\psfig{file=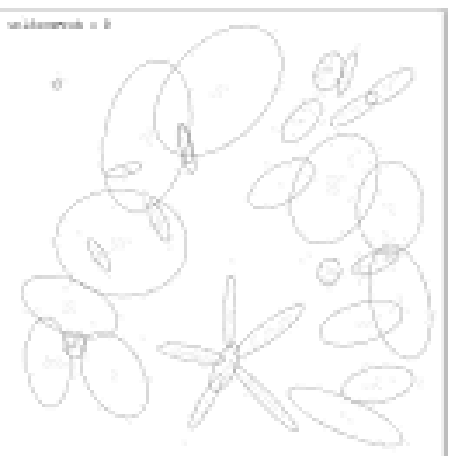}}
\caption{The
Gaussian comprising the mixture model identified from the data
using a search for $k = 1 \ldots 60$ and the BIC criterion.
\label{bic-seq-gauss}
}
\end{figure}

\begin{figure}[htb]
\centerline{\psfig{file=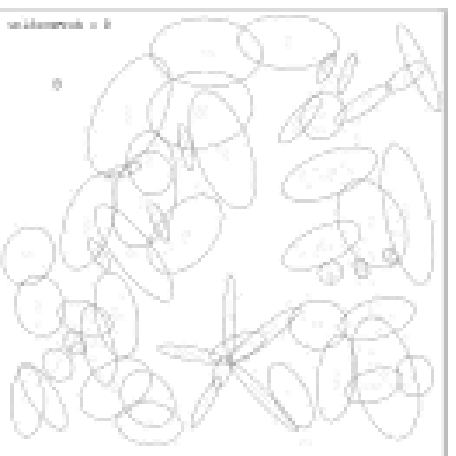}}
\caption{The
Gaussian comprising the mixture model identified from the data
using a search for $k = 1 \ldots 60$ and the AIC criterion.
\label{aic-seq-gauss}
}
\end{figure}

\begin{figure}[htb]
\centerline{\psfig{file=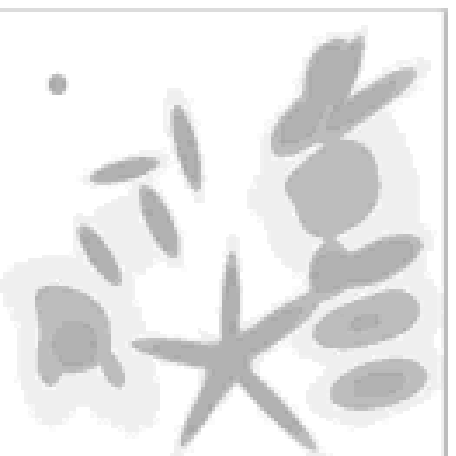}}
\caption{The
PDF Associated with the best BIC model of Figure~\ref{bic-seq-gauss}.
\label{bic-seq-pdf}
}
\end{figure}

\begin{figure}[htb]
\centerline{\psfig{file=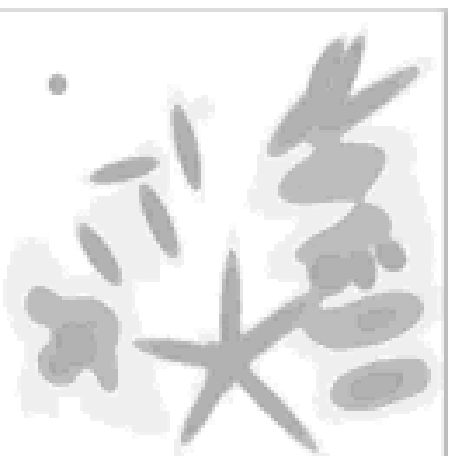}}
\caption{The
PDF Associated with the best AIC model of Figure~\ref{aic-seq-gauss}.
\label{aic-seq-pdf}
}
\end{figure}

\begin{figure}
\centerline{\psfig{file=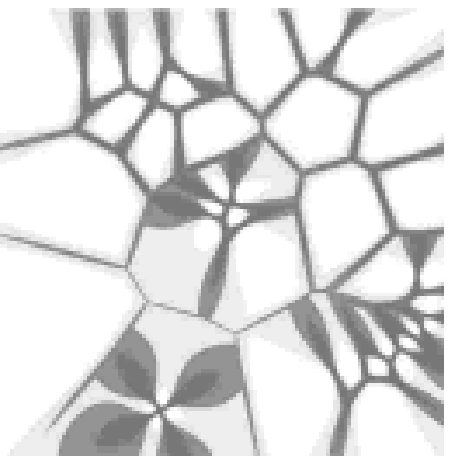,width=0.45\textwidth}\hfil\psfig{file=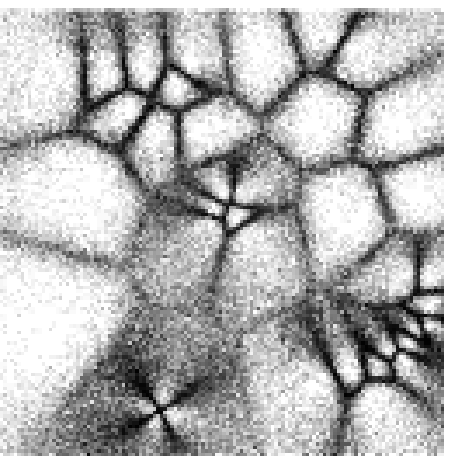,width=0.45\textwidth}}
\centerline{\psfig{file=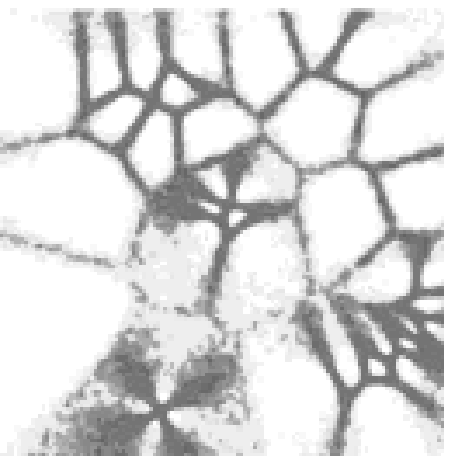,width=0.45\textwidth}\hfil\psfig{file=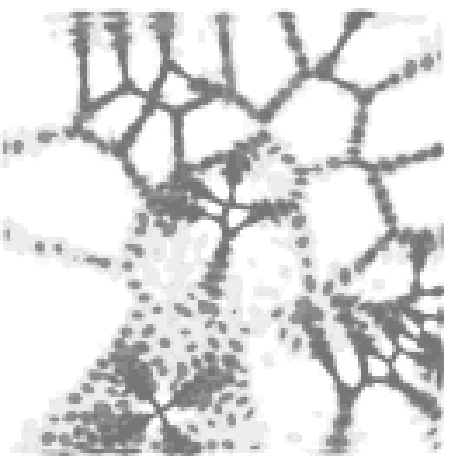 ,width=0.45\textwidth}}
\caption{A comparison of the fixed kernel and EM density estimation 
techniques. The top left panel shows the simulated density
distribution drawn from a Voronoi Tessellation. The top right panel
shows a data set of 100,000 data points drawn from the Voronoi
distribution. The results of density estimation are given in the lower
two panels with the left panel showing a fixed kernel density
estimation (where the kernel has been ``hand tuned'' to provide the
most accurate representation of the data -- see text) and the right
panel the density derived from the EM algorithm using AIC scoring. The
fixed-kernel and EM algorithms both reproduce accurately the broad
features within the simulated data. The EM algorithm does, however,
adapt to the sharpest features in the data whereas the fixed kernel
over smooths these regions. 
\label{compare}
}
\end{figure}

\begin{figure}
\centerline{\psfig{figure=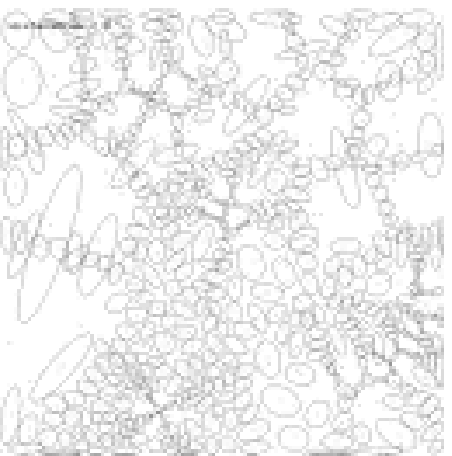,width=0.45\textwidth}\hfil\psfig{figure=connolly.fig11.eps,width=0.45\textwidth}}
\centerline{\psfig{figure=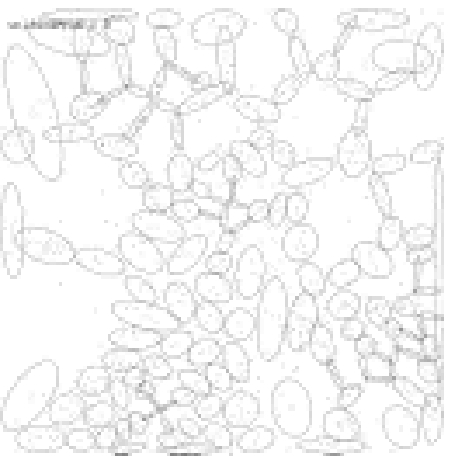,width=0.45\textwidth}\hfil\psfig{figure=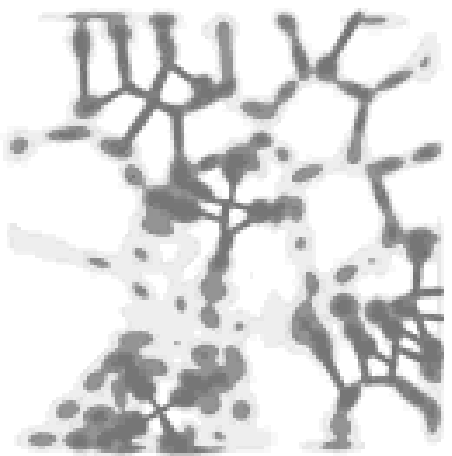,width=0.45\textwidth}}
\caption{A comparison of the density estimation using the AIC and  
BIC scoring criteria when applied to the Voronoi simulation. The top
two panels show the distribution of Gaussians (left) produced by EM
using the AIC score and the resulting density distribution
(right). The lower two panels show the same distributions but for the
BIC scoring criterion. The AIC scoring clearly results in a larger
number of Gaussians which better represent the small scale features
within the data (but with a corresponding loss in the compactness of
the description of the data). In comparison the BIC scoring results in
many fewer Gaussians but does not trace the fine structure as
accurately.
\label{compare1}
}
\end{figure}

\begin{figure}

\centerline{\psfig{figure=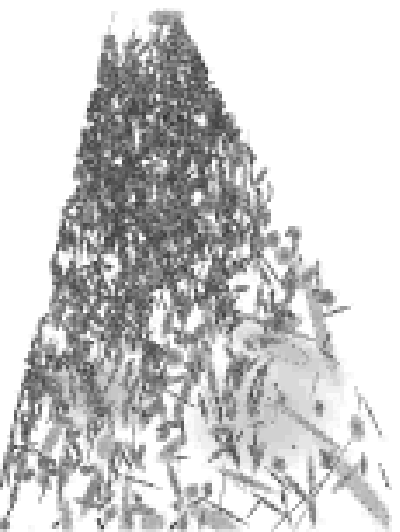,width=0.45\textwidth}\hfil\psfig{figure=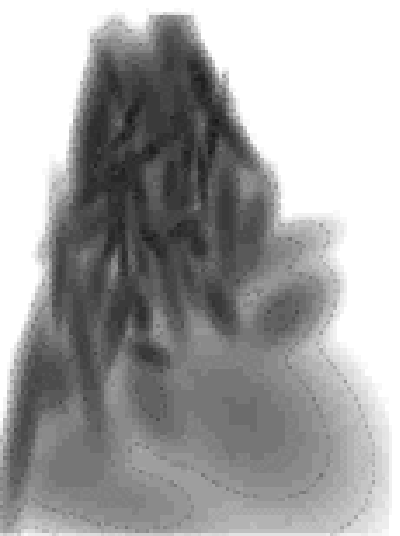,width=0.45\textwidth}}
\centerline{\psfig{figure=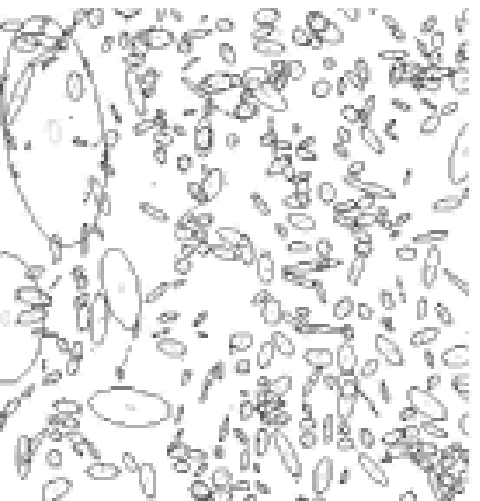,width=0.45\textwidth}\hfil\psfig{figure=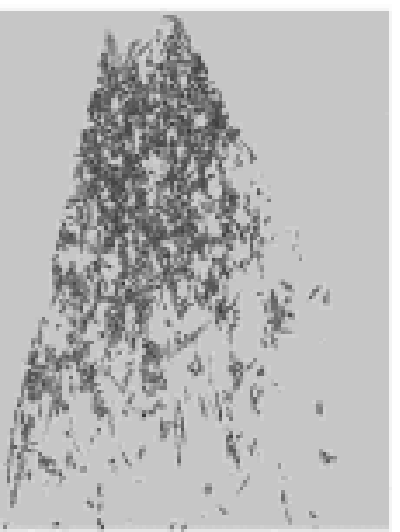,width=0.45\textwidth}\hfil}
\caption{The EM probability density maps (normalized to an integrated 
probability of unity) for the LCRS data discussed in the text. The
top--left panel is for AIC scoring while the top--right panel is for
BIC scoring. The bottom--left panel is a small subset of the data plus
fitted Gaussians taken from the AIC scoring run presented in full in
the top--left panel. The bottom-right panel shows the AIC scoring run
but with a uniform background component in the mixture model. All
three runs used the same splitting probability (0.125) and were run
for 2 hours on a 600MHz PentiumIII with 128 MBytes of memory (although
as illustrated in Figure \ref{time}, EM converges to a solution in
less than 20 minutes).
\label{density}
}
\label{LCRS}  
\end{figure}
 
\begin{figure}

\centerline{\psfig{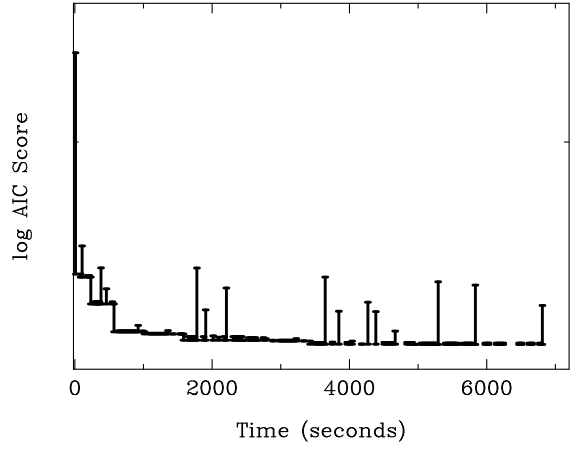}}
\caption{The EM AIC score for the LCRS data as a function of
time (seconds). Each point is presented as an error bar where
the top of the bar represents the AIC after one splitting 
iteration, while the bottom is the beginning AIC score.
\label{time}
}
\end{figure}

\begin{figure}
\centerline{\psfig{figure=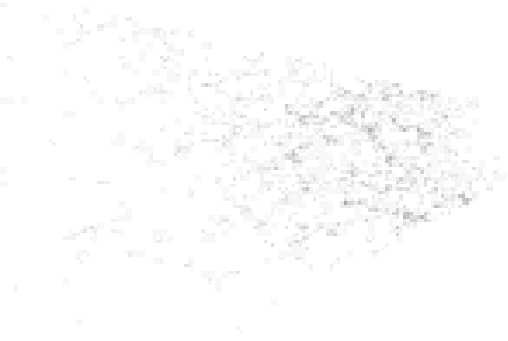,angle=270,width=0.45\textwidth}}
\centerline{\psfig{figure=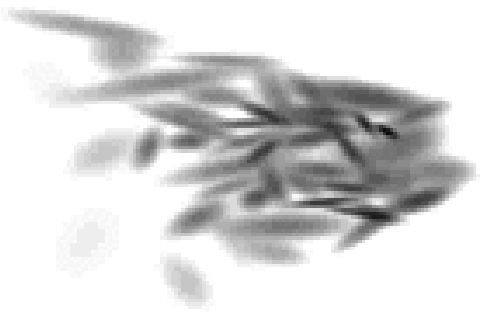,angle=270,width=0.45\textwidth}}
\caption{A thresholded map of EM AIC (top) and BIC (bottom).
\label{threshold}
}
\end{figure}

\begin{figure}

\centerline{\psfig{figure=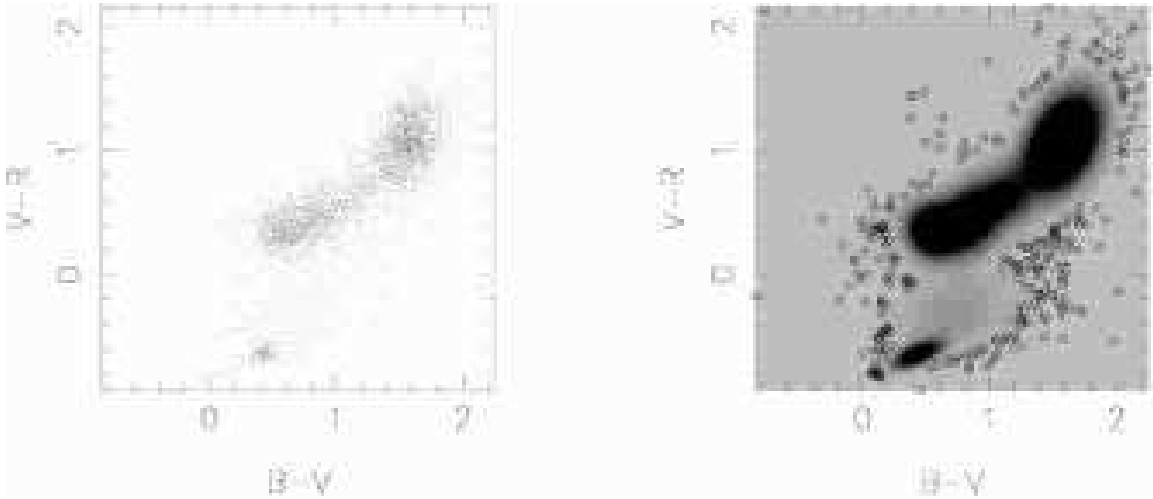,angle=-90,width=6in}}
\caption{The application of EM to d
ensity estimation in color space. The 
left panel shows the distribution of $B-V$ and $V-R$ colors of stellar
sources. The right hand panel shows the density distribution, in
grayscale, of these sources derived from the EM algorithm (using the
AIC scoring criterion) together with the 5\% of the data that is
least likely to be drawn from this distribution (open circles). 
\label{color}
}
\end{figure}

\end{document}

%% file: connolly.tab1.tex
\begin{table}
\begin{center}
\caption{Comparison of AIC and BIC scoring algorithms}
\begin{tabular}{lll}
\tableline\tableline
Scoring Criterion                    & BIC & AIC \\
\tableline
Num Gaussians in Best Model          & 33  & 54  \\
KL-Divergence to true distribution   & 0.207    & 0.067 \\
Seconds to compute using traditional EM & {\bf XXX ?} &  {\bf XXX ?} \\
Seconds to compute using MRKD-trees & 450 & 450 \\
Drawing of Gaussian in best model & Figure~\ref{bic-seq-gauss} &
                                    Figure~\ref{aic-seq-gauss} \\
Drawing of PDF of  best model & Figure~\ref{bic-seq-pdf} &
                               Figure~\ref{aic-seq-pdf} \\
\tableline
\end{tabular}
\end{center}
\end{table}